\documentclass[journal,twoside]{IEEEtran}
\usepackage{cite}
\usepackage{amsmath,amssymb,amsfonts}
\usepackage{algorithmic}
\usepackage{graphicx}
\usepackage{textcomp}
\usepackage{fancyhdr}
\usepackage{booktabs}
\usepackage{multirow}
\usepackage{xcolor}
\usepackage{url}
\def\BibTeX{{\rm B\kern-.05em{\sc i\kern-.025em b}\kern-.08em
    T\kern-.1667em\lower.7ex\hbox{E}\kern-.125emX}}

\begin{document}

\title{Time Distribution Principle Using Measured Traveling Waves in Power Grid}

\author{Haozong Wang, Yayu Yang, Jiahui Yang, Muhamman Umar Afzaal, Yuru Wu, Yan Wen, \\Zhengfa Zhang, and Yilu Liu, \IEEEmembership{Fellow, IEEE}}
\maketitle

\begin{abstract}
Accurate time synchronization is essential for distributed systems. Conventional methods, such as satellite-based synchronization and communication-based approaches, face challenges including signal vulnerability and dependence on communication delay symmetry. This paper proposes a novel time synchronization principle based on traveling wave (TW) measurements in power grids. By leveraging the inherent symmetry of forward and backward TW propagation, the proposed method achieves high-precision time distribution without relying on external time references. The methodology is validated through electromagnetic transient simulations on a modified IEEE 14-bus system. The results demonstrate that under normal conditions, the proposed approach achieves microsecond-level synchronization accuracy. These findings suggest that the TW-based time synchronization principle is a potential alternative to traditional methods, offering improved security and reduced dependence on communication quality.
\end{abstract}


\section{Introduction}

\IEEEPARstart{T}{ime}
 synchronization is a critical requirement in many distributed systems, ensuring data consistency and coordinated operations. High-precision time synchronization technology has been widely applied in various fields, including synchronized measurements in power grids \cite{phadke2008synchronized}, internet synchronization for computer networks \cite{mills1991internet}, and high-frequency trading and transaction in financial systems \cite{angel2015equity}.

A high-precision time synchronization system typically involves two key components. One is the clock source, which provides an accurate time reference. The other is the time distribution system, responsible for delivering the reference time to clients. Currently, advanced high-precision time synchronization technologies can be broadly categorized into two types: satellite-based time synchronization, which relies on global navigation satellite systems (GNSSs), and communication-based synchronization, which utilizes network or dedicated communication channels.

\begin{figure}[tb]
	\includegraphics[width=\linewidth]{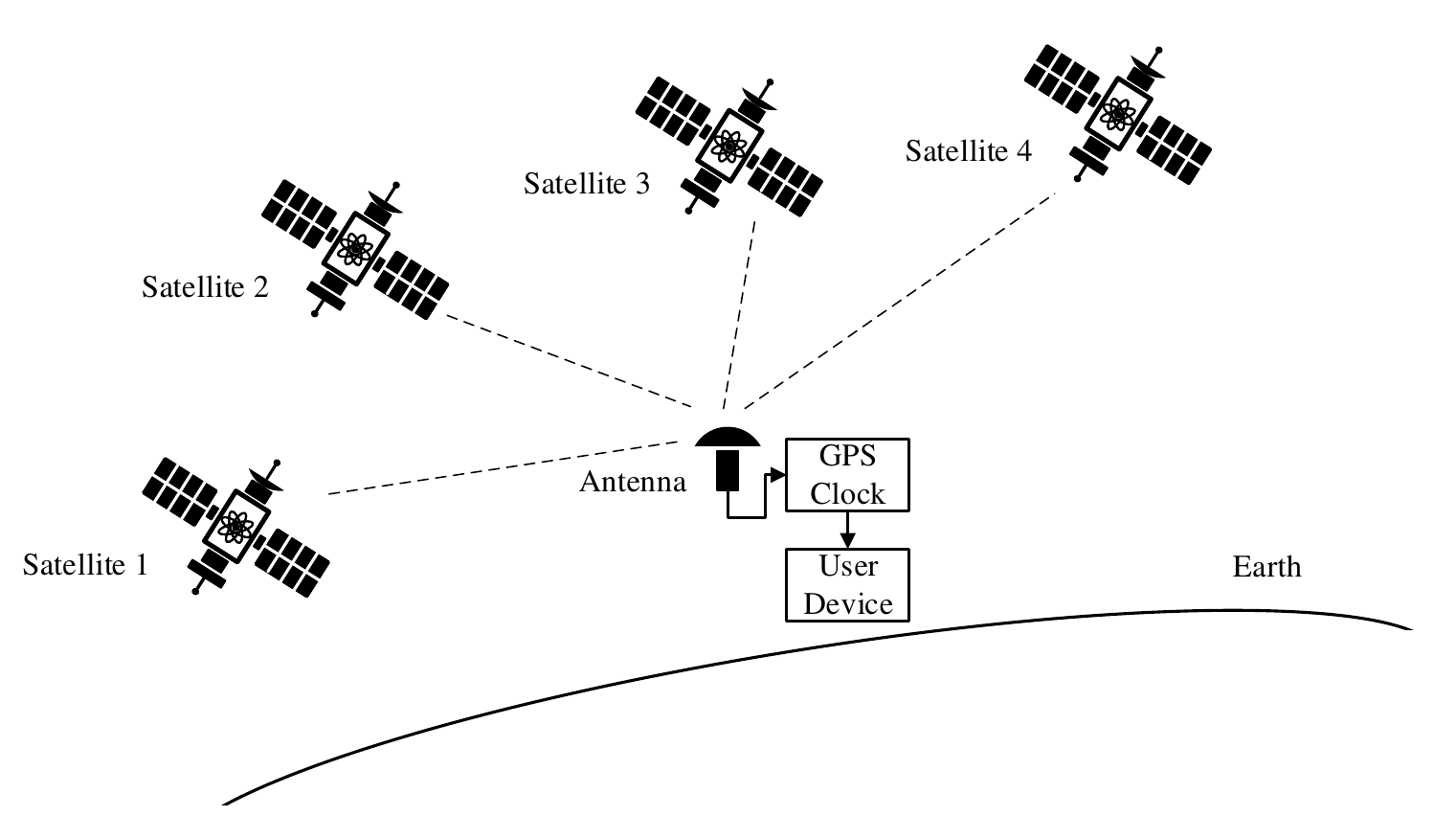}
	\caption{Schematic diagram of satellite-based synchronization \label{fig:GPS}}
\end{figure}

A schematic diagram of satellite-based synchronization is shown in Fig. \ref{fig:GPS}. In this approach, the clock source is the high-precision atomic clocks onboard the satellites. The primary synchronization principle involves real-time calculation of the distance between the satellite and the user's antenna. Under stable signal conditions, this method can achieve a synchronization accuracy of less than 1 $\mu s$. The most widely-used GNSS is the global positioning system (GPS) \cite{lombardi2001time}, which employs the IRIG-B protocol for time distribution \cite{irig_standard_2016}. The IRIG-B protocol allows the usage of cables or optical fibers to transmit time signals from the GPS clock to the client. This process introduces some errors, primarily due to the signal propagation delay in cables or optical fibers. However, since the GPS clock and the client are typically located very close, these errors are minimal \cite{behrendt2006perfect}. For users, the advantages of satellite-based synchronization include simplicity of implementation and high accuracy. However, a significant drawback is its vulnerability to signal loss issues \cite{yao2016impact} or spoofing attacks \cite{shepard2012evaluation}, which can compromise reliability.

\begin{figure}[tb]
	\includegraphics[width=\linewidth]{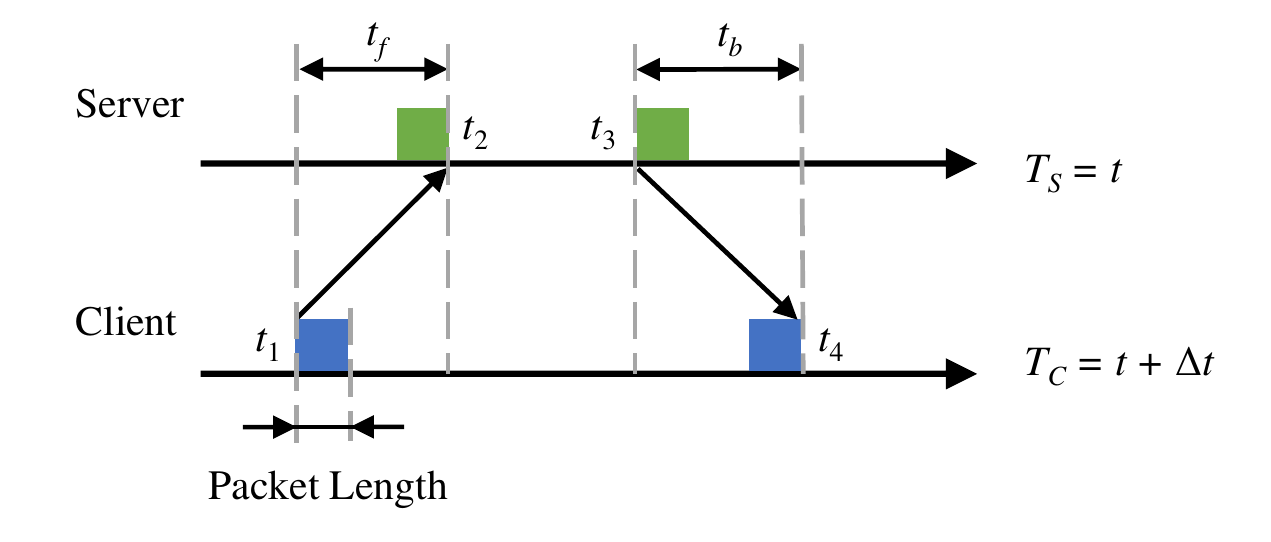}
	\caption{Basic principle of communication-based synchronization 
		\label{fig:communication}}
\end{figure}

Communication-based time synchronization technology refers to utilizing modern communication techniques to distribute the server's time (considered as the accurate clock source) to the client. In this process, the critical technical focus lies in the precise estimation of communication delay. Assuming there exists a time offset $\Delta t$ between the client's clock $T_C$ and the server's clock $T_S$, through a two-way communication process, as shown in Fig. \ref{fig:communication}, both the client and the server record the timestamps  $t_1$ to $t_4$ for sending and receiving packets. If the forward communication time $t_f$ and the backward communication time $t_b$ are constant and equal, $\Delta t$can be calculated using equation \eqref{eq:communication}.

\begin{equation}
	\Delta t = \frac{(t_2-t_1) + (t_3-t_4)}{2}
	\label{eq:communication}
\end{equation}

However, the forward ant backward communication delays are not always symmetric and may vary in time \cite{johannessen2004time}. To minimize this uncertainty, a series of works have modeled communication delays as random variables and employed statistical methods \cite{mills1985network, maroti2004flooding, sommer2009gradient, lenzen2009optimal}, resulting a synchronization accuracy ranging from a few microseconds to tens of milliseconds, depending on the specific communication conditions. For higher precision synchronization requirements, the IEEE 1588 protocol proposes the use of hardware timestamping, which eliminates the impact of software-level jitter, enabling synchronization accuracy of up to 1 microsecond \cite{eidson2002ieee}. Further, the White Rabbit project \cite{serrano2013white, lipinski2011white} employs hardware-based phase detection to replace timestamping, thereby enhancing time resolution. Additionally, it introduces the quantification and calibration of physical medium asymmetry, enabling sub-nanosecond level time synchronization. To summarize, communication-based time synchronization is highly dependent on the quality of both communication and hardware. Achieving high-precision time synchronization requires significantly increased costs.

In this paper, we propose a novel time synchronization principle based on traveling wave (TW) measurement in power grids, which is fundamentally different from the aforementioned two approaches. As is well known, power grid is an artificially constructed network of vast spatial dimensions, where electrical quantities propagate along power lines in the form of TWs. The motivation of this work stems from the fact that the propagation speed and time of TWs are strictly governed by the physical laws of the power system, and therefore, measuring and analyzing TW signals offers a promising approach to achieving time synchronization. Compared to the two well-established technologies mentioned above, the proposed principle provides a different technological perspective with potential advantages: 1) It does not rely on broadcast satellite signals, thereby ensuring higher security; 2) While communication is still necessary, the synchronization accuracy is completely independent of the quality of communication (e.g., symmetry and stability of two one-way communication delays), making it a cost-effective alternative. This paper validates the proposed principle through electromagnetic transient simulations. The results demonstrate that, even under conditions with notable measurement errors, the synchronization accuracy can still reach the microsecond level.

The remainder of this paper is organized as follow. Section \ref{section-2} presents the problem statement. Section \ref{section-3} outlines the proposed methodology, including a introduction of TW model and the time distribution algorithm. Section \ref{section-4} provides simulation-based validation of the proposed principle. Finally, Section \ref{section-5} concludes the paper.

\section{Problem Statement}
\label{section-2}

This section defines the problem by presenting the power grid model and the time distribution system model.

\begin{itemize}
	\item	Power Grid Model
	\vspace{5pt}
	
	Consider a power grid consisting of $n$ nodes and $b$ branches. In this grid, voltages at each of the nodes and currents at each of the branches are assumed to be measurable (that is, $n$ voltage signals and $2b$ current signals). Additionally, communication is assumed to be feasible between the nodes.
	\vspace{5pt}
	
	\item	Time Distribution System Model
	\vspace{5pt}
	
	Consider a distributed system requiring synchronization. Within this system, there is one server capable of providing accurate time, with its clock denoted as $T_S$, and multiple clients whose local clocks, denoted as $T_C^j$, need to be calibrated, where $j$ is the index of clients. Each client clock $T_C^j$ has an error denoted as $\Delta t_C^j$, as expressed in equation \eqref{eq:synchronizationModel}.
	
	\begin{equation}
		\left\{
		\begin{aligned}
			T_S &= t \\
			T_C^j &= t + \Delta t_C^j
		\end{aligned}
		\right.
		\label{eq:synchronizationModel}
	\end{equation}
	
\end{itemize}

For generality, it is assumed that the server and each client in this system correspond to specific nodes in the power grid. This assumption is typically valid, as most distributed systems rely on electrical power for their operation. Therefore, the clients could be indexed as $j = 1, 2, ..., n-1$. At each node, the measured voltage and current data are timestamped by the respective clocks. Consequently, the problem of time distribution can be described as determining each $\Delta t_C^j$ based on the grid measurement data and their associated timestamps.

\section{Methodology}
\label{section-3}

\subsection{The Physical Description of TW Propagation along the Power Line}

\begin{figure}[tb]
	\includegraphics[width=\linewidth]{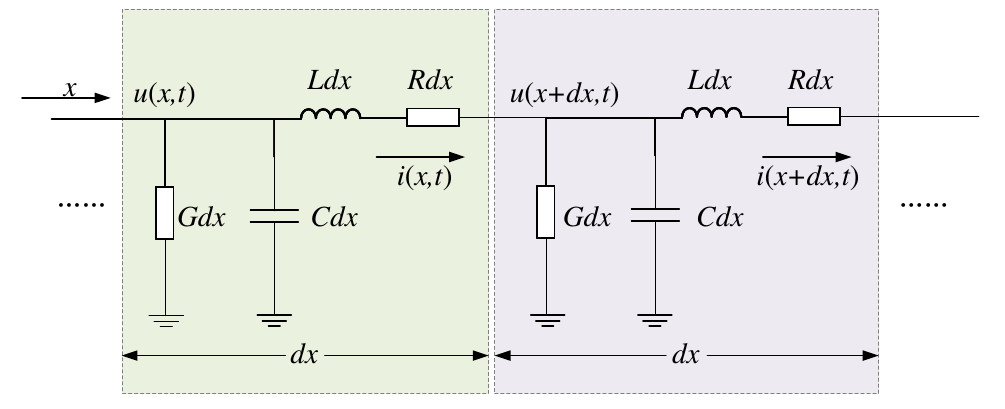}
	\caption{Distributed parameter line model} 
	\label{fig:line}
\end{figure}

Due to the typically long spatial distances of power lines, their accurate description often requires the distributed parameter line model \cite{dommel1969digital}, as shown in Fig. \ref{fig:line}.

Based on Fig. \ref{fig:line}, the telegraph equation could be formulated as:

\begin{equation}
	\left\{
	\begin{aligned}
		-\frac{\partial u(x,t)}{\partial x} &= L\frac{\partial i(x,t)}{\partial t} + Ri(x, t)\\
		-\frac{\partial i(x,t)}{\partial x} &= C\frac{\partial u(x,t)}{\partial t} + Gu(x, t)
	\end{aligned}
	\right.
	\label{eq:telegraph}
\end{equation}

Equation \eqref{eq:telegraph} could be solved in complex frequency domain \cite{marti1982accurate}, the general solution of voltage and current could be formulated as:

\begin{equation}
	\left\{
	\begin{aligned}
		U(x, \omega) &= A_1(\omega)e^{-\gamma(\omega)x} + A_2(\omega)e^{\gamma(\omega)x} \\ &\overset{\text{def}}{=} U_f(x, \omega) + U_r(x, \omega)\\
		I(x, \omega) &= \left[A_1(\omega)e^{-\gamma(\omega)x} - A_2(\omega)e^{\gamma(\omega)x}\right]/Z_c(\omega) \\ &\overset{\text{def}}{=} I_f(x, \omega) + I_r(x, \omega)
	\end{aligned}
	\right.
	\label{eq:solution}
\end{equation}

where $A_1$ and $A_2$ are twice-differentiable functions determined by the boundary conditions; $Z_c$ and $\gamma$ are characteristic impedance and propagation coefficient, determined by the line parameters:

\begin{equation}
	\left\{
	\begin{aligned}
		\gamma(\omega) &= \sqrt{(R+j\omega L)(G+j\omega C)}\\
		Z_c(\omega) &= \sqrt{(R+j\omega L)/(G+j\omega C)}
	\end{aligned}
	\right.
	\label{eq:parameter}
\end{equation}

According to Equation \eqref{eq:solution}, the mathematical form of the voltage TWs can be expressed as:

\begin{equation}
	\left\{
	\begin{aligned}
		U_f(x, \omega) &= A_1(\omega)e^{-\gamma(\omega)x} \\
		U_r(x, \omega) &= A_2(\omega)e^{\gamma(\omega)x}
	\end{aligned}
	\right.
	\label{eq:voltageTW}
\end{equation}

For a specific line of length $l$, the voltage TWs at its sending end ($x=0$) and receiving end ($x=l$) satisfy the following equation:

\begin{equation}
	\left\{
	\begin{aligned}
		U_f(l, \omega) &= U_f(0, \omega)e^{-\gamma(\omega)l} \\
		U_r(0, \omega) &= U_r(l, \omega)e^{-\gamma(\omega)l}
	\end{aligned}
	\right.
	\label{eq:propagation}
\end{equation}

For a specific frequency component, such as the fundamental frequency $\omega_0$, the propagation coefficient is a complex constant, resulting in:

\begin{equation}
	\left\{
	\begin{aligned}
		U_f(l, \omega_0) &= U_f(0, \omega_0)e^{-\gamma(\omega_0)l} = U_f(0, \omega_0) \times re^{-j\theta} \\
		U_r(0, \omega_0) &= U_r(l, \omega_0)e^{-\gamma(\omega_0)l} = U_r(l, \omega_0) \times re^{-j\theta} 
	\end{aligned}
	\right.
	\label{eq:propagationFF}
\end{equation}

Equation \eqref{eq:propagationFF} describes the propagation characteristics of TWs in the complex frequency domain. It can be observed that after propagating a distance $l$, the TW undergoes a phase shift of $\theta$, which corresponds to a certain propagation delay in the time domain. Notably, the forward TW and the backward TW have the same $\theta$. This symmetry serves as the foundation for time synchronization.

\subsection{Time Distribution Between Two Terminals of a Power Line}
\label{section-3b}

Extracting the phase part of equation \eqref{eq:propagationFF} yields:

\begin{equation}
	\left\{
	\begin{aligned}
		\phi_f(l, \omega_0) &= \phi_f(0, \omega_0) - \theta \\
		\phi_r(0, \omega_0) &= \phi_r(l, \omega_0) - \theta 
	\end{aligned}
	\right.
	\label{eq:propagationPhase}
\end{equation}

where $\phi$ is the phase angle of the voltage TWs. Note that there is an implicit assumption of equation \eqref{eq:propagationPhase} that the phase angles at $x=0$ and $x=l$ are calculated using the same clock. In the case of unsynchronized measurement, assuming that the clock at one terminal of the line is accurate, while the clock at the other terminal differs from it by $\Delta t_C$, equation \eqref{eq:propagationPhase} should be re-arranged as:

\begin{equation}
	\left\{
	\begin{aligned}
		\phi_f(l, \omega_0) &= \phi_f(0, \omega_0) - \theta + \delta \\
		\phi_r(0, \omega_0) &= \phi_r(l, \omega_0) - \theta - \delta
	\end{aligned}
	\right.
	, \delta = 2\pi\frac{\Delta t_C}{T}
	\label{eq:synchroEquation}
\end{equation}

where $T$ is the period of fundamental frequency $\omega_0$. Equation \eqref{eq:synchroEquation} could be easily solved as:

\begin{equation}
	\Delta t_C = \frac{T}{4\pi}\left[\phi_f(l, \omega_0)-\phi_f(0, \omega_0)+\phi_r(l, \omega_0)-\phi_r(0, \omega_0)\right]
	\label{eq:dtsolution}
\end{equation}

\subsection{Time Distribution of a Power Grid}

For a power grid consisting of $n$ nodes and $b$ branches, based on the assumption in section \ref{section-2}, there exists in total $n-1$ unknown $\Delta t_C^j$ to be solved. For each branch, an equation can be established in the form described in section \ref{section-3b}, resulting in a total of $b$ equations. If the power grid is radial, then $b=n-1$, and the equations is well-determined, allowing for direct solution. If the power grid contains loops, then $b>n-1$, making the equations over-determined, in which case the least squares method can be used to obtain a solution.

\section{Simulation Verification}
\label{section-4}

\subsection{Test System}

\begin{figure}[tb]
	\includegraphics[width=\linewidth]{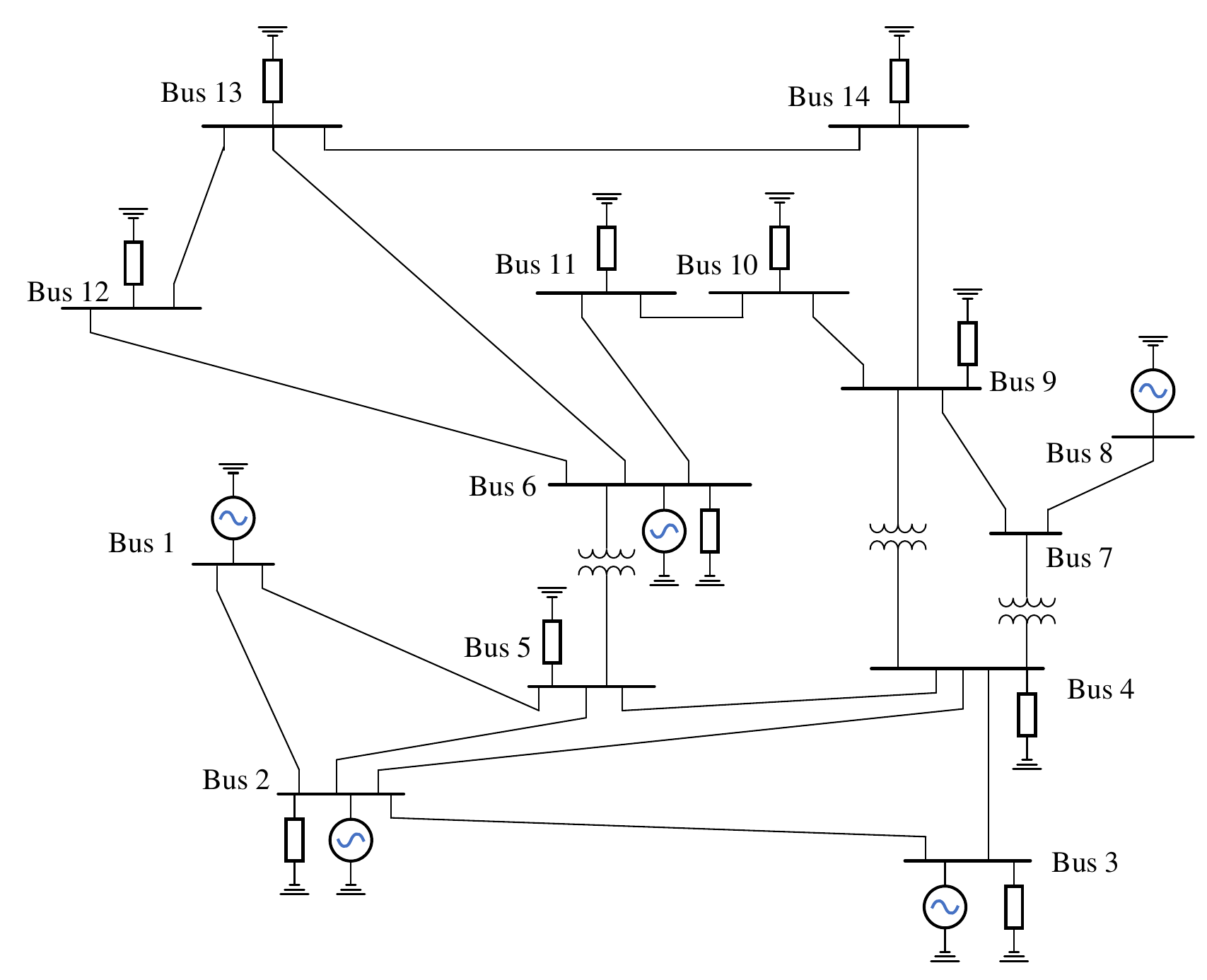}
	\caption{Single-line diagram of the test system} 
	\label{fig:system}
\end{figure}

In this work, the IEEE 14-bus system \cite{pstca14bus} was adopted as the test system, as shown in Fig. \ref{fig:system}. To implement it in PSCAD/EMTDC software, several modifications were made to the original model. In the original model, the lines were represented using lumped parameter impedances. In this study, to simulate the TW transient, the lines were converted to the distributed parameter model. The line lengths were estimated based on the inductive reactance, where a unit inductive reactance of $2.63\times 10^{-3}$ p.u./m was used. For some lines where capacitive reactance data was not provided, the value of $5.81\times 10^{-4}$ p.u./m was adopted as the unit capacitive reactance. Additionally, the grid frequency was adjusted to 50 Hz. The parameters of the model can be found in Table \ref{tab:parameter}.

\begin{table*}[tb]
	\centering
	\caption{Modified IEEE 14-bus system parameter}
	\begin{tabular}{@{}cc|ccc|cccc@{}}
		\toprule
		\multicolumn{2}{c|}{line} & \multicolumn{3}{c|}{\begin{tabular}[c]{@{}c@{}}original lumped\\ parameters\end{tabular}} & \multicolumn{4}{c}{\begin{tabular}[c]{@{}c@{}}modified distributed\\ parameters\end{tabular}} \\ \midrule
		from bus     & to bus     & $R$ (p.u.)                     & $X$ (p.u.)                     & $B$ (p.u.)                    & $R$ (p.u./km)           & $X$ (p.u./km)           & $B$ (p.u./km)           & length (km)           \\ \midrule
		1            & 2          & 0.01938                      & 0.05917                      & 0.0528                      & 8.60E-04              & 2.63E-03              & 2.34E-03              & 22.53667              \\
		1            & 5          & 0.05403                      & 0.22304                      & 0.0492                      & 6.36E-04              & 2.63E-03              & 5.79E-04              & 84.95148              \\
		2            & 3          & 0.04699                      & 0.19797                      & 0.0438                      & 6.23E-04              & 2.63E-03              & 5.81E-04              & 75.40281              \\
		2            & 4          & 0.05811                      & 0.17632                      & 0.034                       & 8.65E-04              & 2.63E-03              & 5.06E-04              & 67.15676              \\
		2            & 5          & 0.05695                      & 0.17388                      & 0.0346                      & 8.60E-04              & 2.63E-03              & 5.22E-04              & 66.22741              \\
		3            & 4          & 0.06701                      & 0.17103                      & 0.0128                      & 1.03E-03              & 2.63E-03              & 1.96E-04              & 65.14191              \\
		4            & 5          & 0.01335                      & 0.04211                      & 0                           & 8.32E-04              & 2.63E-03              & 5.81E-04              & 16.03886              \\
		6            & 11         & 0.09498                      & 0.1989                       & 0                           & 1.25E-03              & 2.63E-03              & 5.81E-04              & 75.75703              \\
		6            & 12         & 0.12291                      & 0.25581                      & 0                           & 1.26E-03              & 2.63E-03              & 5.81E-04              & 97.43291              \\
		6            & 13         & 0.06615                      & 0.13027                      & 0                           & 1.33E-03              & 2.63E-03              & 5.81E-04              & 49.61724              \\
		7            & 8          & 0                            & 0.17615                      & 0                           & 0                     & 2.63E-03              & 5.81E-04              & 67.09201              \\
		7            & 9          & 0                            & 0.11001                      & 0                           & 0                     & 2.63E-03              & 5.81E-04              & 41.90061              \\
		9            & 10         & 0.03181                      & 0.0845                       & 0                           & 9.88E-04              & 2.63E-03              & 5.81E-04              & 32.18436              \\
		9            & 14         & 0.12711                      & 0.27038                      & 0                           & 1.23E-03              & 2.63E-03              & 5.81E-04              & 102.9823              \\
		10           & 11         & 0.08205                      & 0.19207                      & 0                           & 1.12E-03              & 2.63E-03              & 5.81E-04              & 73.15562              \\
		12           & 13         & 0.22092                      & 0.19988                      & 0                           & 2.90E-03              & 2.63E-03              & 5.81E-04              & 76.13029              \\
		13           & 14         & 0.17093                      & 0.34802                      & 0                           & 1.29E-03              & 2.63E-03              & 5.81E-04              & 132.5539              \\ \bottomrule
	\end{tabular}
	\label{tab:parameter}
\end{table*}

The simulation time step is set to 1 MHz, and the sampling rate is set to 10 kHz.

\subsection{Data acquisition}

In the system, all the bus voltages and line currents are measured. According to Section \ref{section-3}, the measured voltage and current signals must be converted into the phases of TW before applying equation \eqref{eq:dtsolution} for time synchronization. This process can be implemented through the following steps:

\begin{itemize}
	\item	Step 1: Convert the time-domain voltage and current signals into phasors. Many PMU algorithms can achieve this process \cite{hostetter2003recursive, friedman1994zero, kamwa2013wide, bi2015dynamic, wu2023optimal}. In this study, the simple Recursive Discrete Fourier Transformation (RDFT) method is adopted.
	
	\item Step 2: Based on equation \eqref{eq:solution} and \eqref{eq:voltageTW}, calculate the TW phasors using the following equation:
	
	\begin{equation}
		\left\{
		\begin{aligned}
			U_f &= \frac{1}{2}(U + IZ_c) \\
			U_r &= \frac{1}{2}(U - IZ_c)
		\end{aligned}
		\right.
		\label{eq:calcTW}
	\end{equation}
	
	where $U$ and $I$ are voltage \& current phasors obtained in Step 1.
	
	\item Step 3:  Extract the phase angles from the resulting TW phasors.
	
\end{itemize}

\subsection{Test result}

\subsubsection{Basic verification}

According to the model described in Section II, we assume that only the time at Bus 2 is accurate, and it is designated as the server. The time measurements at all other buses contain errors and need to be aligned with the time at Bus 2, treating them as clients. To simulate time errors, the measurement data of all other buses are randomly shifted within the range of [-2.5ms, 2.5ms].

Notably, as shown in Fig. \ref{fig:system}, buses 4, 7, and 9 are only connected through transformers. From a practical perspective, this indicates that these three buses are located close to each other and may belong to the same substation. Therefore, we assume that buses 4, 7, and 9 collectively form a single client, sharing the same clock for their measurements. The same assumption applies to buses 5 and 6.

The results of a typical case is shown in Table \ref{tab:result}. As shown in the table, the synchronization error for all clients is less than 0.2 $\mu$s, demonstrating the high accuracy of the proposed method.

\begin{table}[tb]
	\centering
	\caption{Time Synchronization Results}
	\begin{tabular}{@{}cccc@{}}
		\toprule
		Bus & Simulated $\Delta t$ ($\mu$s) & Estimated $\Delta t$ ($\mu$s) & Absolute Error ($\mu$s) \\ \midrule
		1         & 1507            & 1507.17         & 0.170                             \\
		3         & -10             & -10.096         & 0.096                             \\
		4/7/9     & 2004            & 2004.178        & 0.178                             \\
		5/6       & 373             & 372.998         & 0.002                             \\
		8         & 1726            & 1726.165        & 0.165                             \\
		10        & 1193            & 1193.091        & 0.091                             \\
		11        & 430             & 429.993         & 0.007                             \\
		12        & -1266           & -1266.165       & 0.165                             \\
		13        & 832             & 832.057         & 0.057                             \\
		14        & -2083           & -2083.193       & \textcolor{red}{0.193}                             \\ \bottomrule
	\end{tabular}
	\label{tab:result}
\end{table}

To further comprehensively validate the proposed principle, the aforementioned experiment was repeated 1,000 times, and the results are summarized in Table \ref{tab:Replication}.

\begin{table}[tb]
	\centering
	\caption{Replication Test Results}
	\begin{tabular}{@{}ccc@{}}
		\toprule
		Bus & Average Error ($\mu$s) & Max Error ($\mu$s) \\ \midrule
		1         & 0.154            & \textcolor{red}{0.357}        \\
		3         & 0.151            & 0.329        \\
		4/7/9     & \textcolor{red}{0.159}            & 0.304        \\
		5/6       & 0.145            & 0.320        \\
		8         & 0.144            & 0.325        \\
		10        & 0.150            & 0.305        \\
		11        & 0.149            & 0.293        \\
		12        & 0.146            & 0.286        \\
		13        & 0.145            & 0.302        \\
		14        & 0.153            & 0.309        \\ \midrule
		Total     & 0.150            & 0.357        \\ \bottomrule
	\end{tabular}
	\label{tab:Replication}
\end{table}

It could be observed that the proposed method demonstrates stable performance across all replication tests, with an average error of approximately 0.15 $\mu$s and a maximum error not exceeding 0.4 $\mu$s.

\subsubsection{Robustness test}

To verify the robustness of the proposed method, different levels of noise were added to the measured data and tested. The obtained results are summarized in Table \ref{tab:noise}.

\begin{table}[tb]
	\centering
	\caption{Robustness Test Results under Noise}
	\begin{tabular}{@{}cccccc@{}}
		\toprule
		\multirow{2}{*}{Bus} & \multirow{2}{*}{$\Delta t$ ($\mu$s)} & \multicolumn{4}{c}{Absolute Error ($\mu$s)}               \\ \cmidrule(l){3-6} 
		&                        & Clean Signal & 30dB noise & 40dB noise & 70dB noise \\ \midrule
		1                    & 1507                   & 0.170        & 4.248      & 1.381      & 0.236      \\
		3                    & -10                    & 0.096        & 0.768      & 2.802      & 0.028      \\
		4/7/9                & 2004                   & 0.178        & 1.413      & 2.289      & 0.396      \\
		5/6                  & 373                    & 0.002        & 1.159      & 3.689      & 0.24       \\
		8                    & 1726                   & 0.165        & 0.341      & \textcolor{red}{7.200}      & \textcolor{red}{0.517 }     \\
		10                   & 1193                   & 0.091        & \textcolor{red}{37.932}     & 3.795      & 0.317      \\
		11                   & 430                    & 0.007        & 14.744     & 1.246      & 0.263      \\
		12                   & -1266                  & 0.165        & 14.242     & 3.512      & 0.037      \\
		13                   & 832                    & 0.057        & 2.302      & 2.417      & 0.171      \\
		14                   & -2083                  & \textcolor{red}{0.193}        & 9.358      & 2.082      & 0.005      \\ \bottomrule
	\end{tabular}
	\label{tab:noise}
\end{table}

It can be observed that measurement noise affects the accuracy of the proposed principle to some extent. Under 30 dB noise, the maximum error reaches 38 $\mu$s. However, despite this, the method still maintains acceptable performance. With 40 dB noise, the maximum error does not exceed 10 $\mu$s, and under 70 dB noise, which is typically the noise level of sensors in power grids, the proposed principle ensures an error of less than 1 $\mu$s.

\subsection{Discussion}

Based on the above experimental results, the proposed principle can generally ensure a synchronization accuracy of less than 1 microsecond under typical conditions, which meets the requirements of most time synchronization applications. However, it is important to note a key limitation of this method: phase measurements are inherently periodic with a $2\pi$ cycle. This means that the proposed approach assumes all $\Delta t$ values do not exceed the signal period $T$ of the power grid; otherwise, the method will fail.

\section{Conclusion}
\label{section-5}

This paper introduced a novel time synchronization principle based on TW measurements in power grids, fundamentally differing from existing satellite-based and communication-based approaches. By exploiting the inherent symmetry of traveling wave propagation, the proposed method ensures accurate time distribution without reliance on external reference signals. The methodology was validated using simulations on the IEEE 14-bus system, demonstrating that the synchronization accuracy remains below 1 microsecond under normal conditions. The proposed method provides a potential alternative to conventional techniques, offering improved security and reduced dependence on communication quality.

However, one key limitation of this method is its reliance on the assumption that time offsets remain within one power frequency period; exceeding this threshold could lead to synchronization errors. Future research will focus on refining the approach to address this limitation.

\bibliographystyle{IEEEtran}
\bibliography{refs}

@book{phadke2008synchronized,
	title={Synchronized phasor measurements and their applications},
	author={Phadke, Arun G and Thorp, James S},
	volume={1},
	number={2017},
	year={2008},
	publisher={Springer}
}

@article{mills1991internet,
	title={Internet time synchronization: the network time protocol},
	author={Mills, David L},
	journal={IEEE Transactions on communications},
	volume={39},
	number={10},
	pages={1482--1493},
	year={1991},
	publisher={Ieee}
}

@article{angel2015equity,
	title={Equity trading in the 21st century: An update},
	author={Angel, James J and Harris, Lawrence E and Spatt, Chester S},
	journal={The Quarterly Journal of Finance},
	volume={5},
	number={01},
	pages={1550002},
	year={2015},
	publisher={World Scientific}
}

@article{lombardi2001time,
	title={Time and frequency measurements using the global positioning system},
	author={Lombardi, Michael A and Nelson, Lisa M and Novick, Andrew N and Zhang, Victor S and others},
	journal={Cal Lab: International Journal of Metrology},
	volume={8},
	number={3},
	pages={26--33},
	year={2001}
}

@manual{irig_standard_2016,
	title        = {IRIG Serial Time Code Formats},
	author       = {{Telecommunications and Timing Group}},
	year         = {2016},
	note         = {IRIG standard 200-16, archived from the original on 2018-08-26},
	url          = {https://https://www.irigb.com/pdf/wp-irig-200-04.pdf},
	urldate      = {2024-12-30}
}

@article{behrendt2006perfect,
	title={The perfect time: An examination of time-synchronization techniques},
	author={Behrendt, Ken and Fodero, Ken and others},
	journal={Publication, Schweitzer Engineering Laboratories, Inc},
	pages={1--18},
	year={2006},
	publisher={Citeseer}
}

@article{yao2016impact,
	title={Impact of GPS signal loss and its mitigation in power system synchronized measurement devices},
	author={Yao, Wenxuan and Liu, Yong and Zhou, Dao and Pan, Zhuohong and Till, Micah J and Zhao, Jiecheng and Zhu, Lin and Zhan, Lingwei and Tang, Qiu and Liu, Yilu},
	journal={IEEE Transactions on Smart Grid},
	volume={9},
	number={2},
	pages={1141--1149},
	year={2016},
	publisher={IEEE}
}

@article{shepard2012evaluation,
	title={Evaluation of the vulnerability of phasor measurement units to GPS spoofing attacks},
	author={Shepard, Daniel P and Humphreys, Todd E and Fansler, Aaron A},
	journal={International Journal of Critical Infrastructure Protection},
	volume={5},
	number={3-4},
	pages={146--153},
	year={2012},
	publisher={Elsevier}
}

@article{johannessen2004time,
	title={Time synchronization in a local area network},
	author={Johannessen, Svein},
	journal={IEEE control systems Magazine},
	volume={24},
	number={2},
	pages={61--69},
	year={2004},
	publisher={IEEE}
}

@techreport{mills1985network,
	title={Network time protocol (NTP)},
	author={Mills, David L},
	year={1985}
}

@inproceedings{maroti2004flooding,
	title={The flooding time synchronization protocol},
	author={Mar{\'o}ti, Mikl{\'o}s and Kusy, Branislav and Simon, Gyula and L{\'e}deczi, Akos},
	booktitle={Proceedings of the 2nd international conference on Embedded networked sensor systems},
	pages={39--49},
	year={2004}
}

@inproceedings{sommer2009gradient,
	title={Gradient clock synchronization in wireless sensor networks},
	author={Sommer, Philipp and Wattenhofer, Roger},
	booktitle={2009 International Conference on Information Processing in Sensor Networks},
	pages={37--48},
	year={2009},
	organization={IEEE}
}

@inproceedings{lenzen2009optimal,
	title={Optimal clock synchronization in networks},
	author={Lenzen, Christoph and Sommer, Philipp and Wattenhofer, Roger},
	booktitle={Proceedings of the 7th ACM Conference on Embedded Networked Sensor Systems},
	pages={225--238},
	year={2009}
}

@inproceedings{eidson2002ieee,
	title={IEEE-1588™ Standard for a precision clock synchronization protocol for networked measurement and control systems},
	author={Eidson, John C and Fischer, Mike and White, Joe},
	booktitle={Proceedings of the 34th Annual Precise Time and Time Interval Systems and Applications Meeting},
	pages={243--254},
	year={2002}
}

@article{serrano2013white,
	title={The white rabbit project},
	author={Serrano, Javier and Lipinski, M and Wlostowski, T and Gousiou, E and van der Bij, Erik and Cattin, M and Daniluk, G},
	year={2013}
}

@inproceedings{lipinski2011white,
	title={White rabbit: A PTP application for robust sub-nanosecond synchronization},
	author={Lipi{\'n}ski, Maciej and W{\l}ostowski, Tomasz and Serrano, Javier and Alvarez, Pablo},
	booktitle={2011 IEEE International Symposium on Precision Clock Synchronization for Measurement, Control and Communication},
	pages={25--30},
	year={2011},
	organization={IEEE}
}

@article{dommel1969digital,
	title={Digital computer solution of electromagnetic transients in single-and multiphase networks},
	author={Dommel, Hermann W},
	journal={IEEE transactions on power apparatus and systems},
	number={4},
	pages={388--399},
	year={1969},
	publisher={IEEE}
}

@article{marti1982accurate,
	title={Accurate modelling of frequency-dependent transmission lines in electromagnetic transient simulations},
	author={Marti, Jos{\'e} R},
	journal={IEEE Transactions on power apparatus and systems},
	number={1},
	pages={147--157},
	year={1982},
	publisher={IEEE}
}

@article{hostetter2003recursive,
	title={Recursive discrete Fourier transformation},
	author={Hostetter, G},
	journal={IEEE Transactions on Acoustics, Speech, and Signal Processing},
	volume={28},
	number={2},
	pages={184--190},
	year={2003},
	publisher={IEEE}
}

@article{friedman1994zero,
	title={A zero crossing algorithm for the estimation of the frequency of a single sinusoid in white noise},
	author={Friedman, Vladimir},
	journal={IEEE Transactions on Signal Processing},
	volume={42},
	number={6},
	pages={1565--1569},
	year={1994},
	publisher={IEEE}
}

@article{kamwa2013wide,
	title={Wide frequency range adaptive phasor and frequency PMU algorithms},
	author={Kamwa, Innocent and Samantaray, SR and Joos, Geza},
	journal={IEEE Transactions on smart grid},
	volume={5},
	number={2},
	pages={569--579},
	year={2013},
	publisher={IEEE}
}

@article{bi2015dynamic,
	title={Dynamic phasor model-based synchrophasor estimation algorithm for M-class PMU},
	author={Bi, Tianshu and Liu, Hao and Feng, Qian and Qian, Cheng and Liu, Yilu},
	journal={IEEE Transactions on Power Delivery},
	volume={30},
	number={3},
	pages={1162--1171},
	year={2015},
	publisher={IEEE}
}

@article{wu2023optimal,
	title={Optimal PMU design based on sampling model and sensitivity analysis},
	author={Wu, Yuru and Yin, He and Qiu, Wei and Liu, Yilu and Gao, Shengyou},
	journal={International Journal of Electrical Power \& Energy Systems},
	volume={148},
	pages={109004},
	year={2023},
	publisher={Elsevier}
}

@misc{pstca14bus,
	title        = {Power Systems Test Case Archive: 14 Bus System},
	author       = {University of Washington, Department of Electrical and Computer Engineering},
	year         = {2025},
	url          = {https://labs.ece.uw.edu/pstca/pf14/pg_tca14bus.htm},
	note         = {Accessed: 2025-01-15}
}

\end{document}